\documentclass{IEEEtran}
\usepackage{amsmath,amssymb,amsfonts}
\usepackage{graphicx}
\usepackage{color}
\usepackage{multirow}
\usepackage{hyperref}
\usepackage{textcomp,nicefrac}
\usepackage{booktabs}
\usepackage[dvipsnames]{xcolor}
\usepackage{flushend}
\usepackage[numbers,compress]{natbib}
\usepackage{placeins}
\definecolor{orange}{rgb}{1,0.5,0} %James 
\definecolor{purple}{RGB}{128,0,128} % Seda
\definecolor{green}{RGB}{0,128,0} %Christian

\begin{document}

\title{A reconfigurable neural network ASIC for detector front-end data compression at the HL-LHC}

\author{Giuseppe Di Guglielmo, Farah Fahim, Christian Herwig, Manuel Blanco Valentin, Javier Duarte, Cristian Gingu, Philip Harris, James Hirschauer, Martin Kwok, Vladimir Loncar, Yingyi Luo, Llovizna Miranda, Jennifer Ngadiuba, Daniel Noonan, Seda Ogrenci-Memik, Maurizio Pierini, Sioni Summers, Nhan Tran

\thanks{Manuscript received March 26, 2021; (Corresponding e-mail: farah@fnal.gov)}
\thanks{Farah Fahim, Cristian Gingu, Christian Herwig, James Hirschauer, Llovizna Miranda, Nhan Tran are with Fermi National Accelerator Laboratory, Batavia, IL, USA and are supported by Fermi Research Alliance, LLC under Contract No. DE-AC02-07CH11359 with the U.S. Department of Energy (DOE), Office of Science, Office of High Energy Physics}
\thanks{Manuel Blanco Valentin, Farah Fahim, Yingyi Luo, Seda Ogrenci-Memik, Nhan Tran are with Northwestern University, Evanston, IL, USA}
\thanks{Giuseppe Di Guglielmo is with Columbia University, New York, NY, USA}
\thanks{Javier Duarte is with UC San Diego, La Jolla, CA, USA and is supported by the DOE, Office of Science, Office of High Energy Physics Early Career Research program under Award No. DE-SC0021187.}
\thanks{Philip Harris is with Massachusetts Institute of Technology, Cambridge, MA, USA and  is supported by a Massachusetts Institute of Technology University grant.}
\thanks{Martin Kwok is with Brown University, Providence, RI, USA}
\thanks{Jennifer Ngadiuba is with California Institute of Technology, Pasedena, CA, USA}
\thanks{Daniel Noonan is with Florida Institute of Technology, Melbourne, FL, USA}
\thanks{Vladimir Loncar, Maurizio Pierini, Sioni Summers are with CERN, Geneva, Switzerland and are supported by the European Research Council (ERC) under the European Union’s Horizon 2020 research and innovation program (Grant Agreement No. 772369).}
\thanks{Vladimir Loncar is with the Institute of Physics Belgrade, Serbia}
}
\maketitle

\begin{abstract}
%The next generation of high energy particle collider experiments are coping with high luminosity by building detectors with unprecedented granularity necessitating intelligent data processing early in the detector readout chain in order to make most efficient use of available bandwidth.
%The next generation of high-energy particle collider experiments will demand the capability to intelligently process data at unprecedented rates.
%, with increases driven by demand for larger data sets and finely segmented detectors.
%This necessitates efficient transmission of data from the sensors to maximize trigger bandwidth and ultimately the physics program.
%Machine learning algorithms present a promising set of tools to address the associated pattern recognition tasks. Recent progress towards generic real-time inference with neural networks (NNs) has showcased the potential of hardware implementation using FPGAs coupled with high-level synthesis.
%
%Machine learning algorithms present a promising set of tools to address the associated pattern recognition tasks with recent progress towards generic real-time inference with neural networks (NNs) in FPGAs through the use of high-level synthesis (HLS) tools~\cite{cong2011}.
%
Despite advances in the programmable logic capabilities of modern trigger systems, a significant bottleneck remains in the amount of data to be transported from the detector to off-detector logic where trigger decisions are made. We demonstrate that a neural network autoencoder model can be implemented in a radiation tolerant ASIC to perform lossy data compression alleviating the data transmission problem while preserving critical information of the detector energy profile.  
For our application, we consider the high-granularity calorimeter from the CMS experiment at the CERN Large Hadron Collider.
The advantage of the machine learning approach is in the flexibility and configurability of the algorithm. By changing the neural network weights, a unique data compression algorithm can be deployed for each sensor in different detector regions, and changing detector or collider conditions.  
To meet area, performance, and power constraints, we perform a quantization-aware training to create an optimized neural network hardware implementation.  The design is achieved through the use of %of Mentor Graphics Catapult%
high-level synthesis tools and the \texttt{hls4ml} framework, 
%for converting ML models to hardware.%
%The design  
%and has  been  implemented  in  an  LP  CMOS  65\,nm  process. 
and was processed through synthesis and physical layout flows based on a LP CMOS 65\,nm technology node.
The flow anticipates 200\,Mrad of ionizing radiation to select gates, and reports a total area of 3.6$\,\text{mm}^\text{2}$ and consumes 95\,mW of power.
%and is optimized to withstand approximately 200\,Mrad ionizing radiation.  
The  simulated  energy  consumption  per  inference  is 2.4\,nJ.
This is the first radiation tolerant on-detector ASIC implementation of a neural network that has been designed for particle physics applications.
\end{abstract}


%\tableofcontents

% To do list:
% \begin{itemize}
%     \item harmonize charge and energy terminology
% \end{itemize}

\section{Introduction}
\label{sec:introduction}

%Advances in scientific methodologies and theories typically come through breakthroughs in the precision and speed of sensing instrumentation.
\IEEEPARstart{B}{reakthroughs} in the precision and speed of sensing instrumentation are impactful on advances in scientific methodologies and theories.
%commonly present a high degree of dependency on the reliability, sensitivity and quality of the sensing systems used to gather the data required for each designed experiment. 
Thus, a common paradigm across many scientific disciplines in physics has been %such that an increase in the resolution of the sensing equipment should be expected to also bring an increase in either the robustness of the method or the sensitivity of the experiment itself. 
to increase the resolution of the sensing equipment in order to increase either the robustness or the sensitivity of the experiment itself.
This demand for increasingly higher sensitivity in experiments, along with advances in the design of state-of-the-art sensing systems, has resulted in rapidly growing big data pipelines such that transmission of acquired data to the processing unit via conventional methods is no longer feasible. Data transmission is commonly much less efficient than data processing. Therefore, placing data compression and processing as close as possible to data creation while maintaining physics performance is a crucial task in modern physics experiments.

% Hence, the resolution of the system has become nowadays limited not by how fast we are able to gather as much new data as possible, or by the resolution of the sensor itself, but rather by how limited we are to send this data to the processing or storage unit by the transmission line. 

At the CERN Large Hadron Collider (LHC) and its high luminosity upgrade (HL-LHC), extreme collision rates present extreme challenges for data processing and transmission at multiple stages in detector readout and trigger systems. As the initial stage in the data chain, the on-detector (front-end) electronics that read out detector sensors must operate with low latency and low power dissipation in a high radiation environment, necessitating the use of application-specific integrated circuits (ASICs).
In order to mitigate the initial bottleneck of moving data from front-end ASICs to off-detector (back-end) systems based on field-programmable gate arrays (FPGAs), front-end ASICs must provide edge computing resources to efficiently use limited bandwidth through real-time processing and identification of interesting data. Front-end data compression algorithms have historically relied on zero-suppression, threshold-based selection, sorting or summing of data.

Artificial intelligence (AI), and more specifically machine learning (ML), has recently been demonstrated to be a powerful tool for data compression, processing, and analysis in physics~\cite{Albertsson:2018maf,Radovic:2018dip,Bourilkov:2019yoi,Carleo:2019ptp} and many other domains.
While progress has been made towards generic real-time processing through inference including boosted decision trees and neural networks (NNs) using FPGAs in off-detector electronics~\cite{hls4ml,CMSP2L1T}, ML methods have not yet been used to address the significant bottleneck in the transport of data from front-end ASICs to back-end
FPGAs. 

The high-granularity endcap calorimeter (HGCAL)~\cite{CERN-LHCC-2017-023} currently under construction by the CMS experiment~\cite{Collaboration_2008} for eventual use at HL-LHC provides an excellent example of the big data challenges facing high energy physics. %fitting demonstrator for the ASIC ML accelerator technology. 
As an \emph{imaging calorimeter}, the HGCAL includes  %nearly 600\,m$^2$ of active silicon corresponding to
over 6 million readout channels, providing an unprecedented level of segmentation for calorimetry at high-energy colliders. 
In order to provide input to the real-time event filtering (trigger) system of CMS, the HGCAL transmits a stream of trigger data at a frequency of 40\,MHz resulting in massive data rates.  At data creation, two ASICs are used to digitize and %compress
encode trigger data before transmission to back-end
FPGAs for further processing.

% {\color{orange}Because bandwidth constraints prohibit transmission of full resolution data for all channels, HGCAL employs trigger concentrator (ECON-T ) ASICs. At the HL-LHC 40 MHz bunch crossing (BX) frequency for use in the trigger decision, each ASIC selects or compresses 7-bit signals from each of the 48 individual trigger cells producing outputs as low as 48 bits per BX.}

%The next generation of high-energy particle collider experiments will demand the capability to intelligently process data at unprecedented rates.
%, with increases driven by demand for larger data sets and finely segmented detectors.
%This necessitates efficient transmission of data from the sensors to allow for maximal trigger rates. Despite advances in the programmable logic capabilities of modern trigger systems, a significant bottleneck remains in the amount of data to be transported from the front-end to off-detector logic where trigger decisions are made. 

% {\color{red}I think we could put here a few lines on the reason why we cannot simply keep increasing the number of transmission channels/lines, the limitations on this, etc. so that it becomes clear that we need to encode the data to transmit it.}

In this paper, we explore the application of ML algorithms to the task of processing large amounts of data with low latency and low power in a high radiation environment in order to maximize efficient use of limited bandwidth.
% \cite{nilsson2014principles,rahwan2009argumentation}. 
% Recent progress towards generic real-time inference with neural networks (NNs) has showcased the potential of hardware implementation using FPGAs coupled with high-level synthesis (HLS) tools \cite{cong2011high}.
We focus on an ASIC implementation of an autoencoder algorithm that uses a configurable NN to efficiently compress and encode %a large amount of sensed 
data automatically before transmission.  Subsequent stages of data processing can either decode the data or continue analyzing the encoded data.   
In our ASIC implementation, 
the NN architecture is fixed, but exceptional flexibility in application is preserved by making the NN \textit{weights} programmable.
%while minimizing the loss of information due to the encoding process and guaranteeing the required resolution for the gathered data.
We apply our methodology to the specific front-end data transmission challenge of the CMS HGCAL, showing that %We demonstrate that a complex NN model can be optimized for implementation in an ASIC while achieving excellent physics performance using the example of the CMS HGCAL.in addition to performance gains, 
the advantage of our approach lies in the flexibility and configurability of the algorithm, which allows us to generate unique data compression algorithms  depending on HGCAL sensor geometry, sensor location on the detector and the corresponding occupancy and signal patterns, changing accelerator conditions, or changing detector conditions.%, a unique data compression algorithm can be deployed for each sensor, different detector regions, and changing accelerator conditions.  
%The footprint of this chip is near $4\times5$\,mm$^2$, with the area available for additional compression logic measuring roughly 2.5\,mm$^2$.
%A single compression algorithm is desirable, to avoid the need for multiple versions of the ASIC; this requires the algorithm to be robust and potentially reconfigurable across a range of detector conditions.

The remainder of this paper is organized as follows. %First, we discuss related methodologies which enable ASIC neural network implementations. 
In Section~\ref{sec:our_method}, we introduce the HGCAL challenge in greater detail and outline our conceptual approach. Then, in Section~\ref{sec:AE}, we elaborate on the design and training of the autoencoder NN for the specific case of the CMS HGCAL detector. In Section~\ref{sec:cad_tools}, we present the digital implementation of the trained NN in the ASIC. Finally, we summarize our work and discuss future directions in Section~\ref{sec:conclusions}.

\section{System Constraints and Concept}
\label{sec:our_method}

% {\color{blue}The High-Granularity End-cap Calorimeter (HGCAL) in the CMS experiment is composed by over 6 million readout channels, each one of which generates high resolution data at a frequency of 40MHz.} Since the data from all these channels cannot be efficiently transmitted off-detector, the ECON-T ASIC data concentrator chip receives data from 48 channels each with $22b$ fixed point precision while transmitting between $64b$ to $160b$. 

The HGCAL is a major upgrade of the CMS endcap calorimeter planned to for the HL-LHC and provides a fitting demonstrator for the ASIC ML accelerator technology. The HGCAL is described in detail in Ref.~\cite{CERN-LHCC-2017-023}; relevant implementation details that have changed since publication of Ref.~\cite{CERN-LHCC-2017-023} are updated in this article.

\begin{figure}[!t]
    \centering
    \includegraphics[width=0.95\columnwidth]{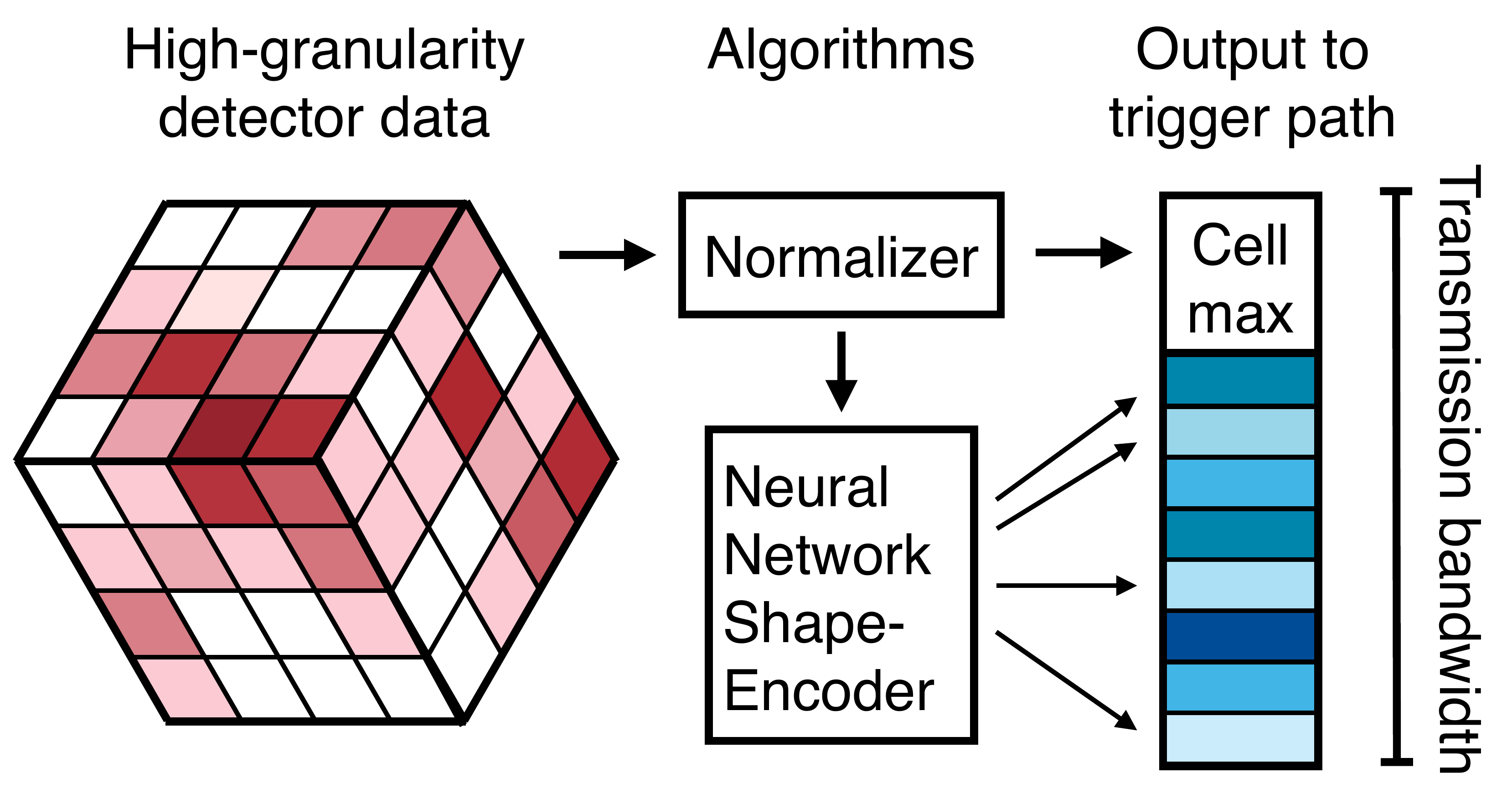}
    \caption{Simplified version of the internal flow of the autoencoder compression task which takes the module energy deposits, normalizes them to the sum of the energy in the module, and then performs shape encoding}
    \label{fig:autoencoder_block}
\end{figure}

This ``imaging calorimeter,'' which includes over 600\,m$^2$ of active silicon and over 6 million readout channels, is composed of $50$ layers of active shower-sampling media interleaved with dense absorber. The active medium of the $28$-layer front electromagnetic compartment is silicon, while the $22$-layer rear hadronic compartment includes both silicon and plastic scintillator. Silicon layers are tiled with $8$'' hexagonal sensor modules, with each module including $48$ logical trigger cells (TC) arranged in three $4\times{}4$ matrices as shown in Fig.~\ref{fig:autoencoder_block}.  While the NN can be configured for both the silicon and scintillator geometries, the silicon geometry is used throughout this paper to illustrate the concepts.

To provide input to the CMS trigger system, data must be transmitted from the on-sensor analog-to-digital ASICs to the all-FPGA back-end detector electronics system at the nominal HL-LHC collision rate of 40\,MHz. Because bandwidth constraints prohibit transmission of data for all 48 TCs at 40\,MHz, a front-end concentrator ASIC (ECON-T) is being developed to compress a single sensor's information before transmission to the back-end trigger electronics. 
%on radiation tolerant optical links.
% For triggering purposes, a single module consists of 48 individual trigger channels in a hexagonal sensor which is illustrated on the left of Fig.~\ref{fig:autoencoder_block}.
Each sensor module produces 7 bits of floating-point charge data for each of the 48 TCs at 40\,MHz. Thus, the lossy compression task of the ECON-T ASIC is to aggregate the 48 7-bit signals from a sensor and compress the data into a range spanning 48--144 bits while maximally preserving the energy pattern of the sensor. The range of the output bits depends on the location of the sensor module in the detector and the number of links available for a given ECON-T ASIC to transmit the data. The number of links allocated will roughly correspond to the average sensor occupancy, which varies by two orders of magnitude over all sensor locations. The exact task depends on handling of data framing and TC address information as well as on whether the ECON-T algorithm operates with fixed or variable latency.  The ECON-T design provides the user a choice among three expert algorithms for TC compression including TC threshold application, sorting and selection of highest energy TC, and aggregation of adjacent TCs.  Unused algorithms are clock-gated to conserve power.

%Therefore, the ECON-T compression task is to preserve as much of the energy pattern in sensor as possible starting with the 1,056 bits of information ($22b \times 48$~channels) and outputting 64$b$-160$b$ of information.  
%{\color{red} @Jim are you happy with this characterization of ECON-T task?}

The ECON-T ASIC is being developed for the LPCMOS (Low Power CMOS) 65\,nm feature size technology and is under active development for CMS.  Because it is located on-detector in a high radiation environment, its design also requires tolerance to single event effects.  
The allocated footprint for the fabrication of this chip is approximately $5\times5$\,mm$^2$. 
%Part of this area is reserved for been reserved for the primary threshold and aggregation expert compression algorithms.
It is expected that sufficient area will be available for potential inclusion of our NN compression logic with an approximate area of $4$\,mm$^2$.
%A single compression algorithm is desirable, to avoid the need for multiple versions of the ASIC; however, multiple algorithms can exist on the ECON-T ASIC and can be configured to choose a single one of the available algorithms at run-time.
The constraints for the compression algorithms are that they should accept new input data at 40\,MHz and complete processing in 50\,ns.
The power budget of the task is less than 100\,mW.

Our contribution is a NN to perform the ECON-T compression task.  
It is a central tenet of our design that the compression algorithm is \textit{reconfigurable}.
Because we are implementing a design for an ASIC, the architecture of the NN will be fixed, but the \textit{weights} of the NN need to be configurable such that the algorithm itself is adaptable.  
This has several advantages. Through reconfiguration, we will be able to: 
\begin{itemize}
    \item \textit{enable} more computationally complex compression algorithms, which could improve overall physics performance or allow more flexible algorithms;
    \item \textit{customize} the compression algorithm of each sensor based on their location within the detector;
    \item \textit{adapt} the compression algorithm for changing detector and collider conditions (for example, if the detector loses a channel or has a noisy channel it can be accounted for or if the collider has more pileup than expected, the algorithm can be adjusted to deal with new or unexpected conditions without catastrophic failure).
\end{itemize}
% this requires the algorithm to be robust and potentially reconfigurable across a range of detector conditions.

For our compression algorithm, we choose to utilize an autoencoder architecture. 
It provides a generic and flexible compression solution, consisting of two NNs: an encoder and a decoder.
The encoding network maps inputs to an intermediate \textit{latent} representation with lower dimensionality than the space of inputs, after which the decoding network aims to recover the original signal.
In the HGCAL application, the encoding NN would compress HGCAL data on the ASIC before transmission to the calorimeter trigger FPGAs for subsequent decoding.
Ultimately, in a final realistic system, we do not anticipate using a full autoencoder architecture because FPGA resources on the back-end FPGA system will not be sufficient to do a full decoding for every sensor.
However, in the absence of understanding how best to use the latent representation later in the processing chain, we optimize performance for an autoencoder because it is a reasonable proxy for the encoder NN encapsulating the salient sensor features such that the image can be decoded from the latent representation.  
Finally, in Fig.~\ref{fig:autoencoder_block}, the compression task is split into two parts: an overall normalization over the entire sensor to preserve the total energy in the sensor and the NN \textit{shape} encoder, which encodes the energy pattern across the sensor.

For the automated design tool flow, it is very important to have a rapid co-design loop between the NN algorithm training and the implementation in hardware in order to understand whether the algorithm is meeting system constraints for power, area, and performance simultaneously.  To achieve this, we use \texttt{hls4ml}~\cite{hls4ml} which translates NNs trained in common open-source ML software frameworks into RTL using high-level synthesis (HLS) tools~\cite{cong2011high}.  The efficacy of this approach will be described in greater detail in the following section.

\section{Algorithm Design and Performance}
\label{sec:AE}

Our task is to design an algorithm that will reproduce the energy pattern in the sensor while simultaneously adhering to hardware constraints, i.e., fitting in the available area within the ECON-T ASIC chip while complying with system latency and power constraints.  Because we are training the algorithm based on a single sensor's energy pattern, we will not be able to optimize for multi-sensor physics performance, such as particle energy resolution.  Ultimately, the physics performance may determine the final system optimization, however, it is beyond the scope of this study.  Therefore, our target is to design an algorithm that reproduces the original sensor energy pattern as accurately as possible through the autoencoder compression-and-decompression bottleneck.  

There are a number of elements needed to design our compression algorithm: a sample of events for training and validation, a preprocessing and normalization block, an optimized NN architecture, and metrics for evaluating the NN performance, both for determining the training loss and the final network evaluation.
An essential aspect of the training procedure is quantization-aware training (QAT), i.e., we approximate bit-accurate reduced precision for all of the NN calculations during training.  QAT is known to be much more performant than post-training quantization (PTQ), where the training is done using 32-bit floating-point operations, which are then truncated post-training to fixed-point or integer representations. In a previous study of the  \texttt{QKeras}~\cite{qkeras} tool, QAT was performed for an LHC trigger task. It was found that with PTQ, the minimum bit width possible without loss of performance was 14 bits while with QAT, the same performance could be achieved with 6-bit weights.  Thus, PTQ would lead to more than 4-fold increase in the area of an ASIC design based on the bit operations hardware design metric~\cite{2020arXiv200408906K}.  Therefore, we use \texttt{QKeras} for training the NNs in this study.

\paragraph*{Training Sample}
Test energy patterns in the sensors are simulated using top-quark-pair events overlaid with 200 simultaneous collisions per bunch crossing in the CMS software framework~\cite{cmssw}.
These simulated events create a sample of typical energy patterns in the HGCAL sensors, which we use as a realistic proxy for the sensor data.  
%{\color{red} @Jim is okay with this (minimal) description of the "data".}

\paragraph*{Preprocessing}
The compression task is factorized into normalization and shape extraction components, as illustrated in Fig.~\ref{fig:autoencoder_block}.  The first stage of ECON-T processing for all algorithms is to expand the 7-bit floating-point TC data to the inherent 22-bit fixed-point TC data.
The sum value of all 48 TC is identified and used to normalize the charge distribution across the full sensor (and the sum of TC charge is included in the final data payload to allow subsequent interpretation of normalized TC data).
%Thus, the maximum charge is transmitted directly to back-end while normalized cell values are passed on to the encoding NN representing only the radiation pattern.
The normalized NN inputs are truncated to 8 bits to allow a more compact NN implementation, while ensuring that any omitted cells constitute less than 1\% of the total energy recorded within a module.
% Normalizing NN inputs reduces the required precision from 22 to 8 bits, allowing a significantly smaller design area with negligible impact on the physics performance. 
%\subsection{Tools}

\begin{figure}[!htb]
\centering
\includegraphics[width=0.8\columnwidth]{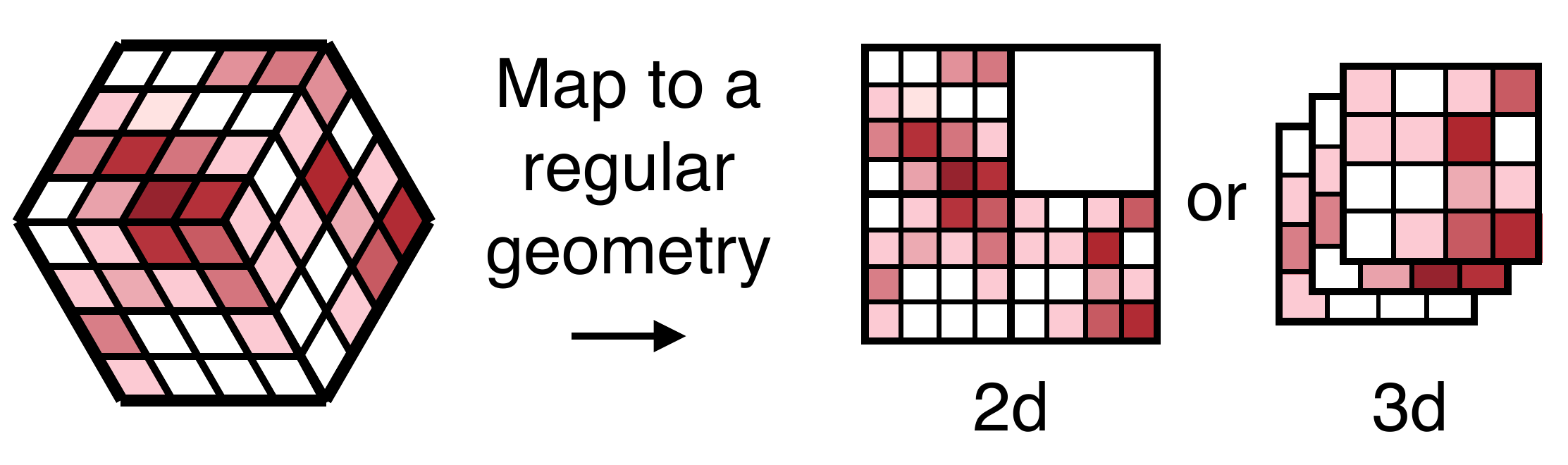}
\caption{Mapping the hexagonal sensor geometry to potential Cartesian representations for convolutional layer operations.} \label{fig:conv}
\end{figure}

\paragraph*{NN Architecture}
The encoding NN architecture consists of successive layers that sequentially process the input data.
%A deep neural network encoder is composed of successive neural network layers, which process the input data. For our HGCAL application, we study convolutional and fully-connected NN layers.
Convolutional layers are used to extract spatial features from images through the application of filters: matrices of configurable parameters.
While convolutional layers use relatively few parameters, convolution requires many multiply-accumulate operations (MACs).
Conversely, a fully-connected layer multiplies a vector of input elements by a matrix of configurable weights, generally requiring more configurable parameters and fewer MACs than a convolution.
The impact of the choice of precision for all internal parameters (constrained by the available area on chip) is accounted for by training inherently quantized models with the \texttt{QKeras} package~\cite{qkeras}.
Because the HGCAL sensor data compression task takes as input an image data representation, we consider a convolutional NN layer as a natural approach.  Typical convolutions rely on the input being in a Cartesian representation, though other shapes can be explored in future work.  Here, we map the hexagonal sensor shape to a more typical Cartesian arrangement as illustrated in Fig.~\ref{fig:conv}, which simplifies the training and hardware implementation.

% Once trained, models are translated to HLS code using the hls4ml~\cite{hls4ml} package, converted to RTL, and the final area is given by place and route.
\paragraph*{Training and Performance Metrics}
The performance of the autoencoder is based on how well the original image is reproduced after encoding and decoding.  
We quantify the difference between raw and decoded HGCAL data with the energy mover's distance (EMD)~\cite{emd}.
Given a particular normalized energy distribution, one may physically relocate some energy fraction $dE$ by a spatial distance $dx$, leading to a new distribution with an associated rearrangement cost $dEdx$.
This notion can be extended to define an ``optimal transport'' between two energy distributions $\mathcal{A}$ and $\mathcal{B}$, as a re-mapping that minimizes this total rearrangement cost, EMD($\mathcal{A},\mathcal{B}$).
%This minimized cost defines the EMD between the two energy distributions.
While the performance of the autoencoder is assessed with EMD, taking into account the complete hexagonal sensor geometry, this metric is not used directly in the algorithm training, as it involves nondifferentiable and computationally intensive operations.
Models are instead trained with a modified $\chi^2$ loss function incorporating cell-to-cell distances, as a fast approximation of EMD.
Specifically, individual TCs are re-summed into all physical groups of approximately $2\times2$ and $3\times3$ ``super-cells'' based on the full hexagonal cells 
with corresponding $\chi^2$ values computed for the coarsened images.
The total loss sums each such $\chi^2$ together, resulting in a comparatively lenient penalty when mis-reconstruction occurs only on small spatial scales.
Including these additional $\chi^2$ terms in the training procedure is found to yield significant improvements to the autoencoder performance, as measured with EMD.
To ensure an unbiased NN optimization, the data are randomly partitioned into separate samples for training (80\%) and validation (20\%), with training termination set by the loss observed in the validation sample.

\begin{figure}[!t]
\centering
\includegraphics[width=0.9\columnwidth]{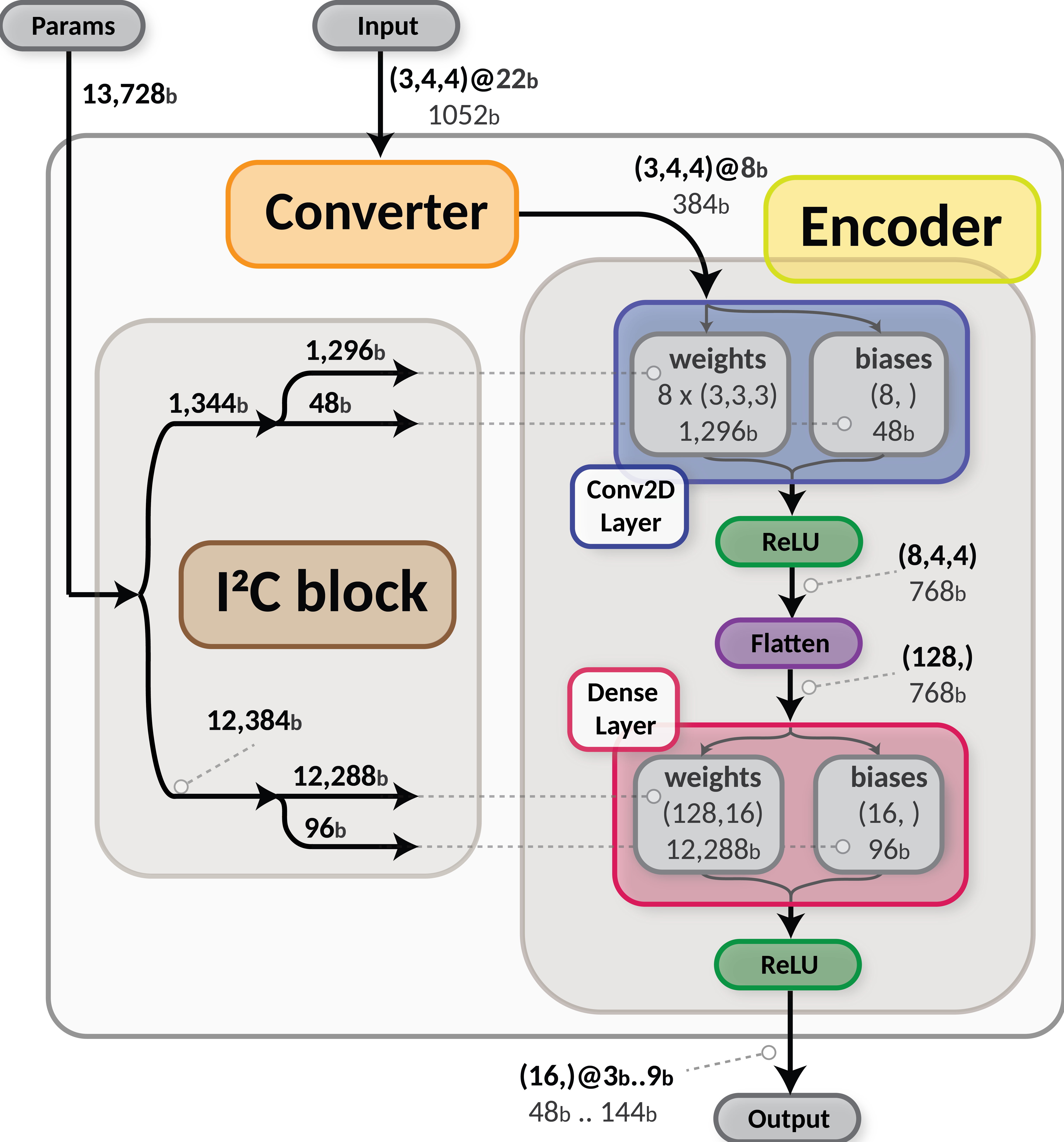}
\caption{The autoencoder neural network architecture and data flow for the baseline encoder model} \label{fig:data_flow}
\end{figure}

%\subsection{Encoder Model Architecture and Optimization}
\subsection*{Baseline Encoder Model}

A simple encoding NN with a single convolutional and dense layer architecture is investigated.
Normalized inputs from hexagonal sensors are arranged into three arrays of $4\times4$ to form a regular geometry.
The convolution layer consists of eight $3\times3\times3$ kernel matrices, giving a $8\times4\times4$ output after convolution.
It was found that more than eight kernel matrices brought negligible performance improvement.  
These 128 values are flattened and fed through a dense layer to yield 16 9-bit output values.
ReLU activations~\cite{relu1,relu2} are applied before and after the dense layer.
This leads to a total of 2,288 weight parameters (dominated by the 2,064 parameters used to configure the dense layer), each of which are specified with 6-bit precision.
A single inference requires a total of 4,448 MACs, with similar requirements from the convolution (2,400) and dense layers (2,048).
The size and complexity of this baseline model are constrained by area, on-chip memory and interfaces, and power, which impose additional optimization considerations.
The encoder architecture with the reconfigurable weights is illustrated in Fig.~\ref{fig:data_flow}
%\todo{The design has been implemented in an Low-Power CMOS 65nm process, takes 615K gates, occupies a total area of 2.5 mm$^2$, consumes 280 mW of power, and optimized to withstand approximately 200Mrad ionizing radiation.
%Also add a sentence on considerations for number of weights as it relates to I2C interface.}

% \vspace{-10pt}

\subsection*{Optimization Considerations and Comparisons}

While the presence of a single convolutional layer is critical for good physics performance of the algorithm (approximated by the EMD between input and decoded images), adding more filters or additional convolutional layers only weakly improves physics performance, at the expense of significantly increased area.
Changes in the number and size of the dense layers yield more dramatic differences.

\begin{figure}[!t]
\centering
\includegraphics[width=0.48\textwidth]{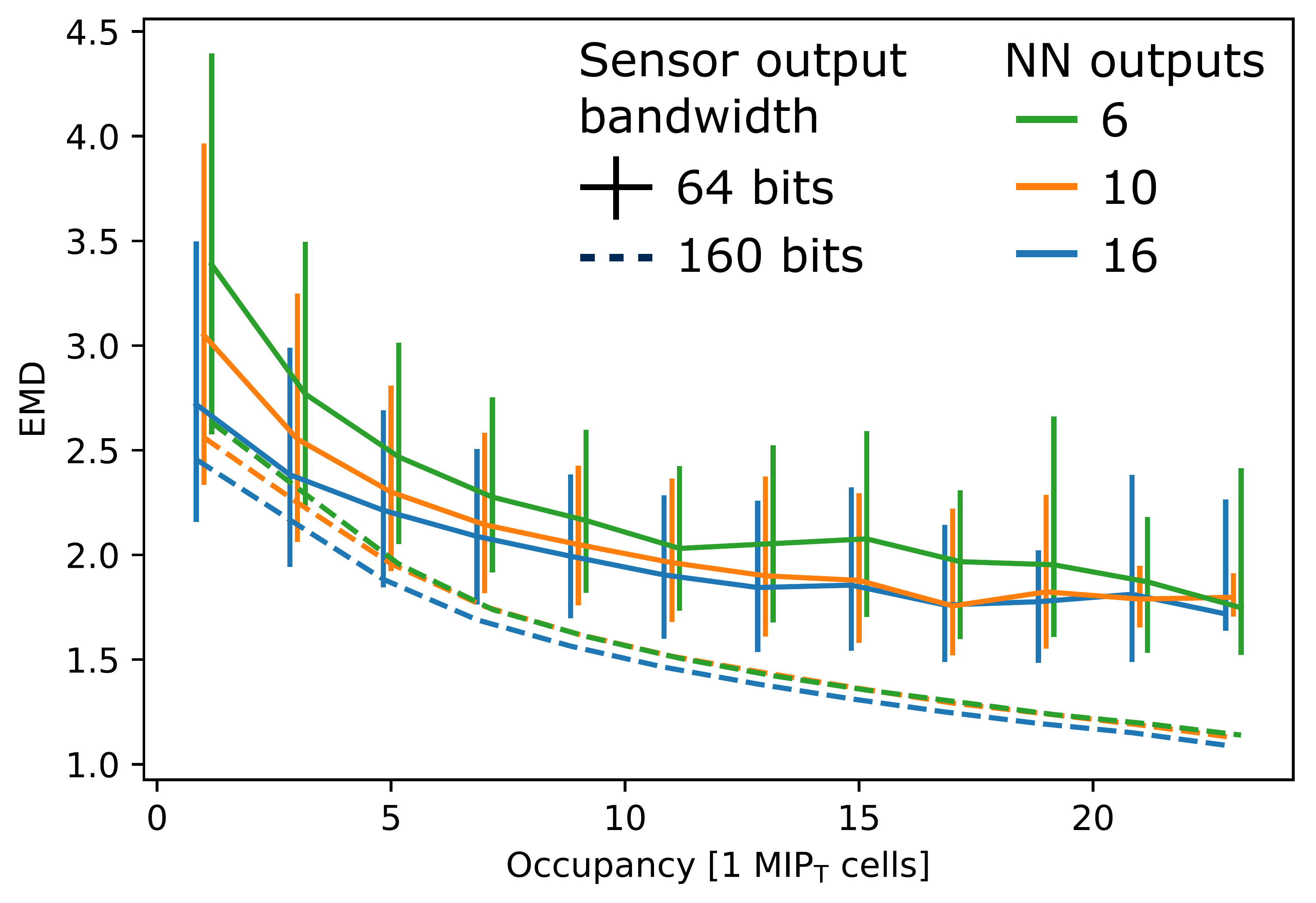}
\caption{Median EMD for decoded HGCAL images from the validation dataset, as function of sensor occupancy for six NN configurations.  Vertical lines (suppressed for the 160-bit configurations) denote 68\% EMD intervals. Occupancy is defined as the number of TCs with signals exceeding one minimum ionizing particle divided by cosh\,$\eta$ where $\eta$ is the pseudorapidity of the TC. (Results shown for version of NN with maximum of 10 bits for each of 16 outputs rather than 9 bits as described in the text.)}
\label{fig2}
\end{figure}

Figure~\ref{fig2} shows a sweep over the number of dense layer outputs, where remaining aspects of the design are fixed based on hardware constraints:
the precision of outputs and weights are coherently varied to ensure that both the total number of outputs and the weight bits are fixed.
Architectures featuring many outputs with lower relative precision consistently outperform their counterparts.
% NN performance can be compared to traditional algorithms such as \textit{thresholding} where low-energy cells are suppressed or \textit{aggregation} where nearby cells are combined.
% Table~\ref{tab1} compares the EMD between HGCAL data and the results of all three compression schemes. 
The autoencoder is robust across a variety of conditions and performs well in the high-occupancy regime, which poses the greatest challenge for trigger reconstruction.

% \begin{table}[!t]
% % increase table row spacing, adjust to taste
% \renewcommand{\arraystretch}{1.3}
%  %if using array.sty, it might be a good idea to tweak the value of
% % \extrarowheight as needed to properly center the text within the cells
% \caption{EMD for NN and traditional compression algorithms.}
% \label{tab1}
% \centering
% \begin{tabular}{r|c|c|c}
% Data set & NN & Aggregation & Thresholding \\
% \hline
% %Inclusive 64b & 2.7\pm0.7 & 0.9\pm0.9 & 0.6\pm1.6 \\
% %High-occ. 64b & & & \\
% %\hline
% Low occupancy & $2.7\pm0.7$ & $1.0\pm0.9$ & $2.1\pm1.5$ (57\% null)\\
% High occupancy & $1.9\pm0.8$ & $2.3\pm0.8$ & $2.2\pm1.5$ (13\% null)\\
% \hline
% \end{tabular}
% \end{table}

\paragraph*{Reconfigurability}
Figure~\ref{fig2} also demonstrates how the same NN encoder can be re-optimized and configured for new data-taking conditions, by comparing sensors in detector regions requiring low- and high-throughput.
The maximum data throughput of 144 bits from 16 9-bit outputs can be reduced through fully configurable selective truncation.  Expected use cases include transmission of (48, 80, 112, 144) bits from 16 (3, 4, 7, 9) bit outputs, though the network can also be configured to transmit fewer than 16 outputs, or a mix of precisions.

%%%%%%%%%%%%%%%%%%%%%%%%%%%%%%%%%%%%%%%%%%%%%%%%%%%%%%%%%%%%%%%%%%%%%%
%%%%%%%%%%%%%%%%%%%%%%%%%%%%%%%%%%%%%%%%%%%%%%%%%%%%%%%%%%%%%%%%%%%%%%
%%%%%%%%%%%%%%%%%%%%%%%%%%%%%%%%%%%%%%%%%%%%%%%%%%%%%%%%%%%%%%%%%%%%%%
%%%%%%%%%%%%%%%%%%%%%%%%%%%%%%%%%%%%%%%%%%%%%%%%%%%%%%%%%%%%%%%%%%%%%%

\section{Implementation Methodology and Results}
\label{sec:cad_tools}

In this section, we detail the implementation of the trained NN described in Sec.~\ref{sec:AE} in the ECON-T ASIC. We discuss the design and verification flow, the architectural and design exploration, steps required for deployment in a radiation environment, design performance metrics, and finally the implementation results.
%\dots
%steps required to go from a trained \texttt{QKeras} model to a synthesizable IP block validated for the ECON-T ASIC.

%\textcolor{red}{
%We may organize this discussion in 4 subsections:
%\begin{itemize}
%\item{Technology Node Choices and Tool Flow}
%\item{Architectural Optimizations and DSE}
%\item{Digital Implementation in Radhard Environment}
%\item{Performance Metrics}
%\end{itemize}
%}

%\begin{figure}[!t]
%\centering
%\includegraphics[width=\columnwidth]{imgs/pdf/area_dse}
%\caption{Area costs and resource breakdown for pipelined implementations. From left to right, the II values are 1, 2, and 4.} \label{fig:area_dse}
%\end{figure}

\begin{table}[t!]
    \begin{center}
    \caption{\label{tab:area_dse}Area breakdown for pipelined implementations. The results are from Catapult HLS estimations and areas are in $\mu m^2$.}
    \begin{tabular}{c|r|r|r}
        % \multicolumn{4}{*}{Key parameters} \\
        Initiation Interval & Total Area & Register Area & MUX Area  \\
        \toprule
         1 & 1,138,242	& 925 & 0 \\
         2 & 891,195 & 5,228 & 12,989 \\
         4 & 765,877 & 8,503 & 16,089 \\
         8 & 699,988 & 8,509 & 16,252 \\
        \bottomrule
    \end{tabular}
    \end{center}
\end{table}

\subsection*{Algorithm to Accelerator Development}
%Technology Node Choices
For our design flow, we adopted the \texttt{hls4ml} framework~\cite{hls4ml} to automate the mapping of ML models onto reconfigurable logic. For this work, we extended \texttt{hls4ml} to our ASIC flow.
Traditionally, hardware designers utilize hardware description languages (HDLs) and a level of abstraction known as the Register Transfer Level (RTL).
In recent years, HLS has become an alternative for generating hardware modules from code written in programming languages such as C/C++. HLS comes with significant benefits: it raises the level of abstraction and reduces the simulation time; it simplifies the verification phases; and finally, it makes the exploration and evaluation of design alternatives easier.
The original flow of \texttt{hls4ml} generates state-of-the-art synthesizable C++ code and HLS directives from the ML-model specifications. The generated code is then fed into the Vivado HLS tool to generate an accelerator in HDL RTL code for the deployment on Xilinx FPGAs~\cite{vivadohls2020}.
We extended \texttt{hls4ml} to support Mentor's Catapult HLS~\cite{catapulthls2020} tool and target our specific 65\,nm LP CMOS technology for ASIC fabrication.
We integrated the HLS-generated code with a SystemVerilog RTL IP of the programmable I\textsuperscript{2}C peripheral\footnote{The authors use \emph{controller/peripheral} in place of \emph{master/slave} when referring to such I\textsuperscript{2}C devices or processes~\cite{osha_resolution}.}.
%We generated the RTL design by combining the HLS created RTL with a SystemVerilog construct of the programmable i\textsuperscript{2}c peripheral.
%
We finally created a component database and layout to be incorporated into the ECON-T ASIC top-level assembly using a digital implementation flow. The standard flow was modified to accommodate automatic triple modular redundancy implementation for HLS-generated RTL integrated with other SystemVerilog modules.

We complemented our design flow with a robust validation and verification methodology across the various refinement steps.  We validated the C++ HLS code against the \texttt{QKeras} trained model to guarantee the model's functional correctness. Earlier in the design flow, we also performed dynamic and static verification of the synthesizable specifications: we checked design rules with static analysis of the C++ HLS code (Mentor CDesignChecker~\cite{hlsverific2020}), measured coverage metrics (Mentor CCov~\cite{hlsverific2020}), and finally, ran simulation-based equivalence checking. For the HLS-generated RTL code, we followed a more traditional simulation-based verification to ensure optimized power, area, and speed.  

%Several verification steps are undertaken at this stage to identify any failures that might have appeared during compilation, as well as to improve performance. Design rules are checked on the C++ specification by performing static analysis (Mentor CDesignChecker). Then C simulation and code coverage (Mentor CCov) are executed. Finally a C\&RTL co-simulation is performed for equivalence checking. The RTL Verilog code subsequently follows the traditional on-chip digital implementation method undergoing various simulation steps to ensure optimized power, area and speed. The final design IP block is generated by creating a component database and layout to be incorporated in the ECON-T ASIC top level assembly.
%

\subsection*{Architectural Exploration}

\texttt{hls4ml} coupled with the industry standard Catapult HLS (ver. 10.6) tool allowed us to explore the cost and performance trade-offs of various micro-architectural hardware implementations for our ML model. We decided on a pipelined implementation for our accelerator to increase concurrent execution as an early design decision.
A pipelined design can process new inputs every $N$ clock cycles, where $N$ is the initiation interval (II) of the design. Table~\ref{tab:area_dse} shows the area breakdown for different II values (1, 2, 4, 8). It is noticeable that although the area is higher for II$=1$, the required resources are mostly functional logic to implement a highly-parallel datapath, i.e. there are no multiplexers.
A higher II value implies less design parallelism and more functional-resource reuse. This choice reduces the overall area, but the resource breakdown shows an increase in control logic (MUX) and registers. 
%\textbf{which are more sensitive to SEE \dots}
An II of 1 was ultimately selected so that new inputs may be processed in sync with a single clock operating at 40\,MHz LHC crossing frequency.

\begin{figure}[!t]
\centering
\includegraphics[width=\columnwidth]{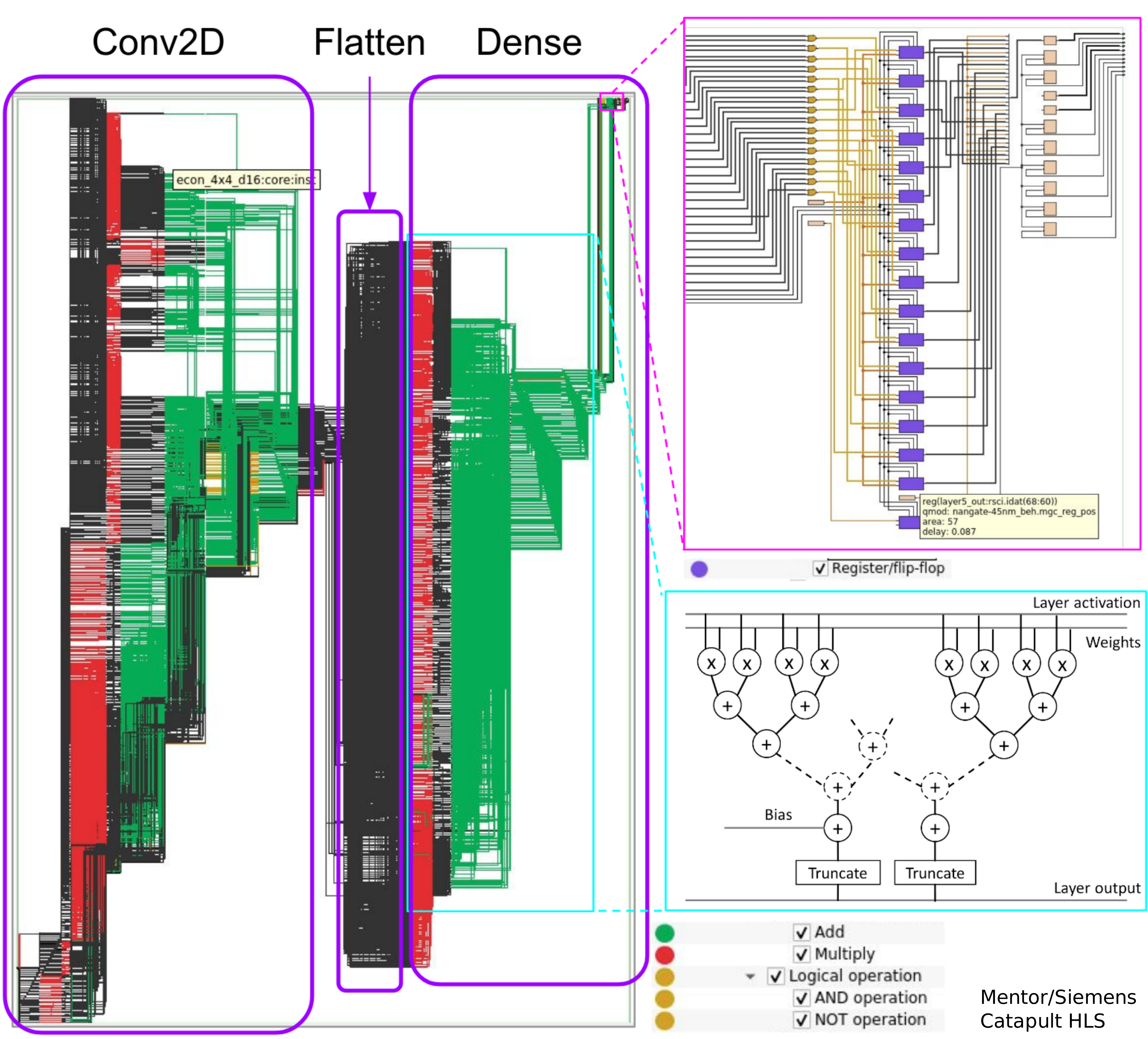}
\caption{Encoder RTL Schematics, the basic structure of the convolution and dense layers can be seen at the schematic level on the left and zoomed in images are provided for the output and MAC portions on the right} \label{fig:rtl}
\end{figure}

We used a fixed-point representation (\texttt{ac\_fixed}~\cite{hlslibs2020}) for the input, intermediate, and output parameters of our ML model designed with \texttt{hls4ml}. This choice provided us with a high degree of flexibility for exploring the area and accuracy trade-off of the ML-model hardware implementations obtained with HLS.  The RTL schematics for the encoder block are shown in Fig.~\ref{fig:rtl}.  The basic structure of the convolution and dense layers can be seen at the schematic level. The top right and bottom right diagrams are zoomed in portions of this schematic depicting the output and MACs. %Fig.~\ref{fig:fxd_dse} shows an example of this exploration.

% \begin{figure}[!t]
% \centering
% \includegraphics[width=0.8\columnwidth]{imgs/pdf/fxd_dse}
% \caption{Fixed-point precision affects the area and latency trade-off. \textcolor{ForestGreen}{THIS CAN BE REMOVED}} \label{fig:fxd_dse}
% \end{figure}

\subsection*{Digital Implementation in a Radiation Environment}
The digital design consists of three major functional blocks: (i) A converter which is a classical module designed with HLS; (ii) An encoder, which uses \texttt{hls4ml}; (iii) and an I\textsuperscript{2}C peripheral which uses a SystemVerilog RTL code. The converter is used for normalizing the 48 (22\,b) inputs to 48 (8\,b). An encoder is used for data classification and further compression to 16 (9\,b) outputs. To have a flexible and reconfigurable algorithm, all the parameters (13,728\,b)  can  be  setup  via  the  I\textsuperscript{2}C  interface  on-chip. The programming of the I\textsuperscript{2}C peripheral takes less than 50\,$\mu$s corresponding to a total of 1,716 I\textsuperscript{2}C clock cycles, utilizing an 8\,b input bus. Once the weights are setup, the algorithm adds a total latency of 2 bunch crossing (BX) cycles to the trigger path---one cycle to convert and another cycle to encode resulting in total inference latency of 50\,ns and a new input accepted every 25\,ns.

\paragraph*{Integrated Converter, Encoder and I\textsuperscript{2}C peripheral}
 
\begin{figure}
    \centering
    \includegraphics[width=.4\textwidth]{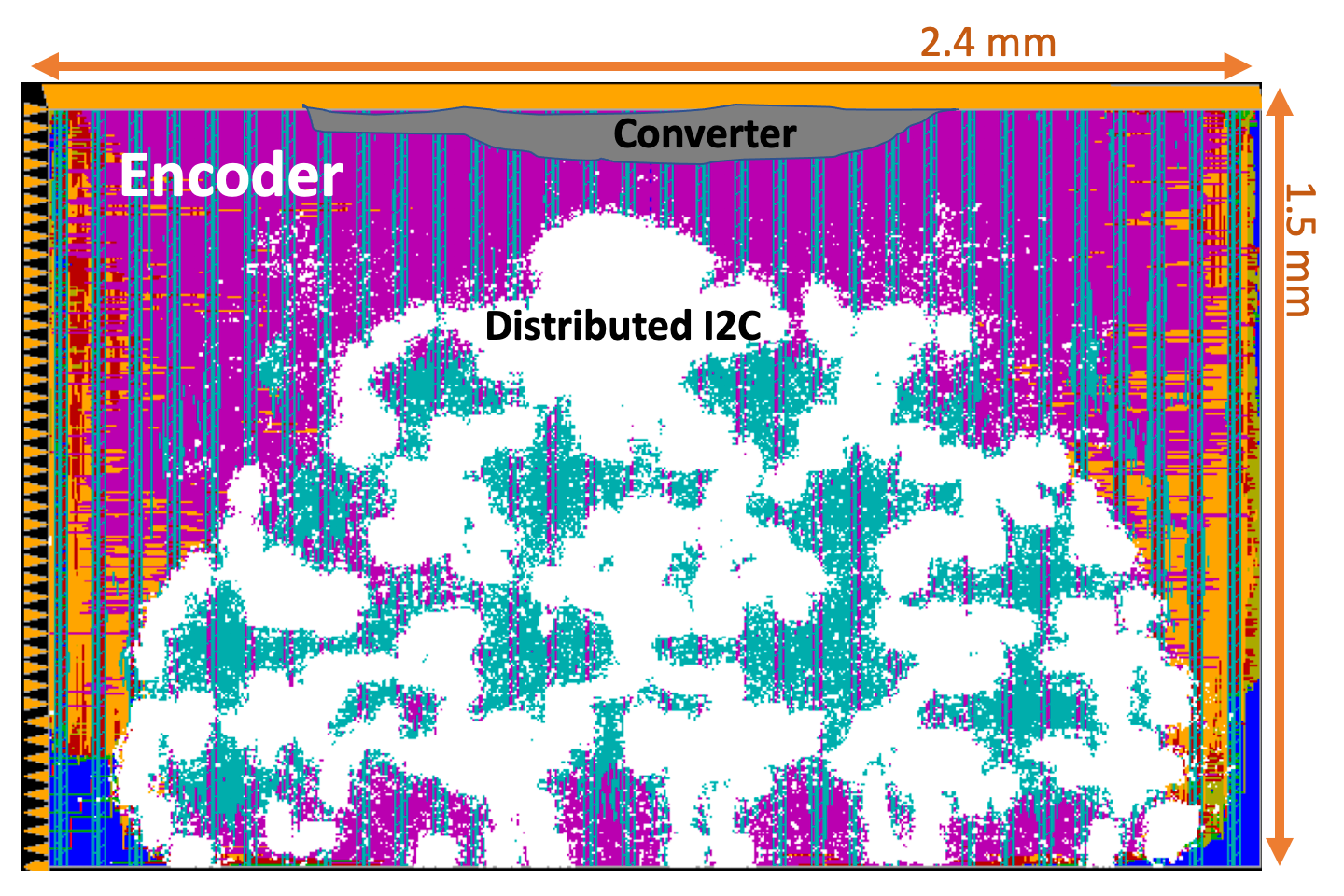}
    \caption{Design floor-plan with an integrated converter, encoder and I\textsuperscript{2}C peripheral occupying a total area of 3.6 mm\textsuperscript{2}. The converter is highlighted in grey, the I\textsuperscript{2}C peripheral in white and the rest of the area is occupied by the encoder.}
    \label{fig:Autoencoder}
\end{figure}

 An integrated approach to the development is needed in order to avoid routing congestion of connecting the weights to the appropriate layers across the encoder. The floor-plan of the digital implementation occupying 2.4\,mm$\times$1.5\,mm is shown in Fig~\ref{fig:Autoencoder}. The converter logic is located near the data input at the top of the design, majority of the area is occupied by the encoder, interleaved with the distributed I\textsuperscript{2}C network.

\paragraph*{Design Considerations for Total Ionizing Dose Performance}

Apart from all requirements considered above, our design must guarantee on-detector circuit reliability in the high radiation environment of HL-LHC~\cite{alia2017lhc,huhtinen1996radiation}. The circuitry should withstand total ionizing dose of approximately 200\,Mrad over the lifetime of the experiment along with high SEE rates~\cite{schrimpf2004radiation,petersen2011single,dodd2010current}. Since previous measurement results have indicated that the average time delay of all cells from the 65\,nm LP process library increases after 200\,Mrad irradiation~\cite{Casas:2275135}, minimum size cells are avoided. Normal Vt standard cell technology library is used. The implementation uses concurrent multi-mode multi-corner static timing analysis for ensuring performance. The foundry worst case libraries are a good stand-in for modelling radiation damage.
%Implementation uses worst case timing libraries to ensure performance after radiation damage. 
All weights are stored in registers and no SRAMs or DICE cells~\cite{8194929} are used.

\begin{figure}[!htb]
    \centering
    \includegraphics[width=.4\textwidth]{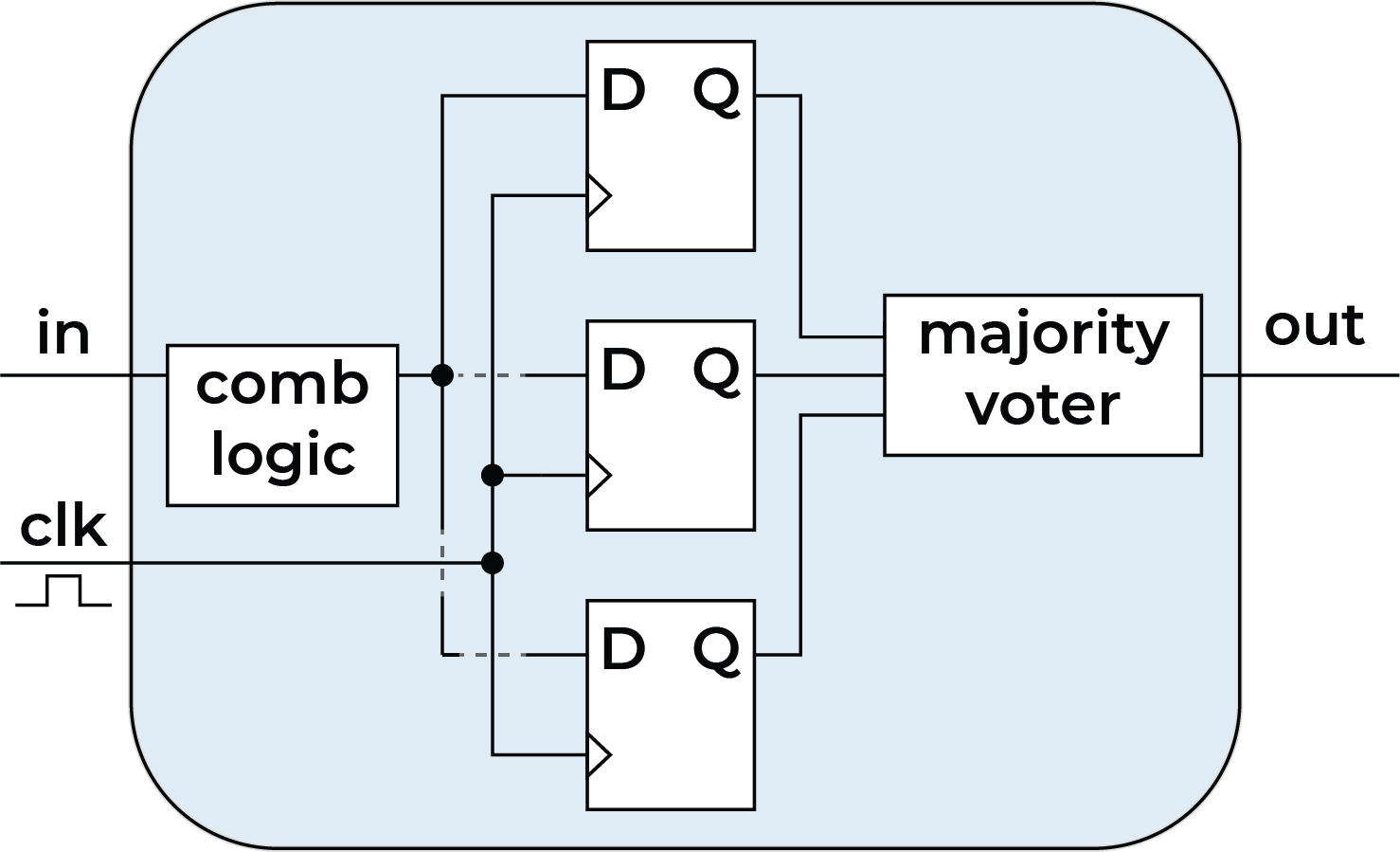} \\
    \caption{Triple modular redundancy scheme used for the encoder and converter. Each register is triplicated and a majority voter determines the output.}
    \label{fig:tmr}
    \vspace{.3cm}
    \includegraphics[width=.4\textwidth]{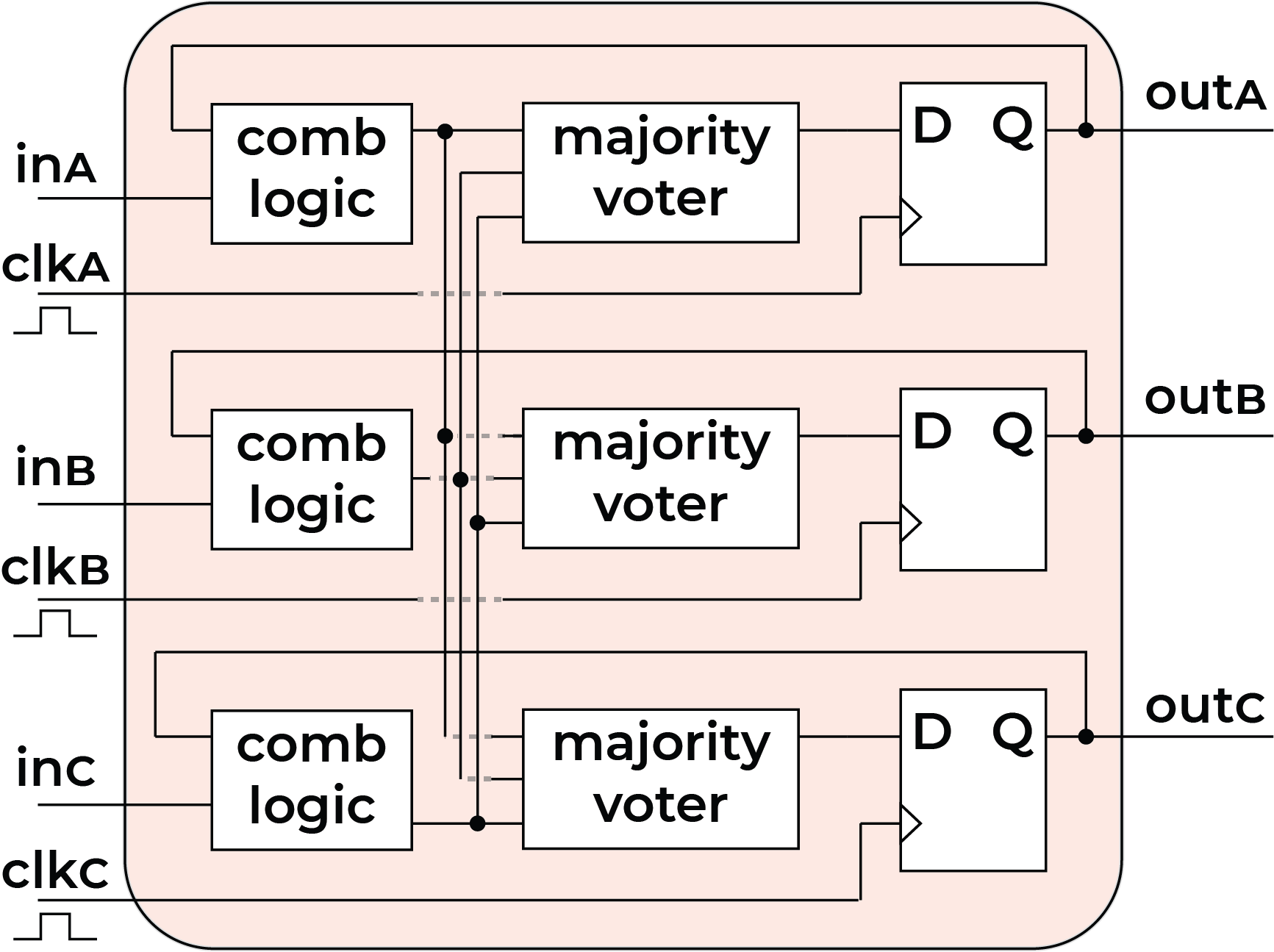}
    \caption{Full module triplication is used for the I\textsuperscript{2}C peripheral. All combinational logic within the module is triplicated, which is used by three majority voters to form the inputs to triplicated registers. Feedback from the output of the registers enables autocorrection and protects against accumulating errors due to single event upsets over time.}
    \label{fig:full_tmr}
\end{figure}

\FloatBarrier
\begin{table*}[ht]
    \centering
    \caption{\label{tab:metrics} Design (D) and verification (V) metrics from model generation to verified IP.}
        \begin{tabular}{p{.3\textwidth}|r|p{.1\textwidth}|p{.2\textwidth}}
        STEP                                 & TIME     & ITER.              & SIZE                                    \\
        \toprule
        Model generation (D)                 & 0.98s    & \multirow{2}{*}{50-100} & \multirow{2}{*}{1089 C++ LoC}           \\
        C Simulation (V)                     & 0.14s    &                         &                                         \\
        \hline
        High-level synthesis (D)             & 00:30:17 & \multirow{2}{*}{3-100}    & \multirow{2}{*}{39,716 Verilog LoC}     \\
        RTL Simulation (V)                   & 00:00:46 &                         &                                         \\
        \hline
        Logic synthesis (D)                  & 06:04:19 & \multirow{5}{*}{6}      & \multirow{2}{*}{769,481 gates}          \\
        Gate-level simulation (V)            & 00:25:19 &                         &                                         \\
        Place and route (D)                  & 39:33:11 &                         & \multirow{3}{*}{806,255 gates}        \\
        Post-layout simulation (V)           & 00:51:41 &                         &                                         \\
        Post-layout parasitic simulation (V) & 01:51:30 &                         &                                         \\
        SEE simulation (V)                  & 04:17:00 &
            &   \\
        \hline
        Layout (D)                           & $\sim{}$00:20:00 & \multirow{2}{*}{1}      &  \\
        LVS \& DRC (V)                       & $\sim{}$01:00:00 &                         &         \\ 
        \bottomrule
        \end{tabular}
\end{table*}
\FloatBarrier

\paragraph*{Single Event Effect Mitigation}

%Specifically, ionizing radiation can cause the electrons of silicon circuits to build-up charge over a certain amount of time which might, eventually, cause a false trigger event in the component itself. The output of digital stages in electronic circuits are usually considered to have specific meaning regarding some part of the global information processed by the circuit itself. This means that false triggers altering the output of these stages can cause a chain reaction that could potentially lead to hazardous and/or invalid results.

%These effects are commonly known as \emph{Single Event Effects} (SEE). 
Mitigating single-event effects (SEEs) is a critical step in the ASIC implementation for effective performance in the HL-LHC environment. Several  %
techniques have been proposed and used over the years to tackle this specific problem~\cite{fulkerson2006single,kuboyama2009single,gong2008new}. 

Triple modular redundancy (TMR) is a well-known technique to protect digital circuits against the undesirable effects of SEEs~\cite{lyons1962use,habinc2002functional}. Depending on the functionality of the block auto-correction features might be required for registers which store data.  
% DO NOT DELETE THESE COMMENTS
%When the probability of radiation affecting more than one of these replicated units within a single block is considered sufficiently low (as is the case considered in this work), TMR has been observed to be effective for mitigating SEE. In our particular application, at any given instant, we expect that at least two of the three replicated blocks are going to remain unaffected by radiation, even if the output of one of them does suffer from SEE. 

We have used two different TMR implementations: simple TMR with triplicated registers and a majority voter for the data path shown in Fig.~\ref{fig:tmr} and fully triplicating the entire module as shown in Fig.~\ref{fig:full_tmr} for the I\textsuperscript{2}C peripheral for storing weights. 

Since new data arrives to the encoder block every $25$\,ns, no auto-correction techniques are required. On the other hand, the values of the weights set by the I\textsuperscript{2}C peripheral (parameters of the NN) are vital as they are central to the vector multiplications used in NNs. Once programmed these are not expected to change over lengthy periods of time, hence, auto correction techniques are used to ensure that register errors due to single event upsets do not accumulate over time. As shown in Figure~\ref{fig:full_tmr}, all combinational logic within the module is triplicated, which is used by three majority voters to  form  the  inputs  to  triplicated  registers.  Feedback  from  the  output  of the registers enables autocorrection. This method does require the I\textsuperscript{2}C peripheral to be clocked periodically.

\subsection*{Performance Metrics and Implementation Results}

Table \ref{tab:metrics} lists metrics characterizing our design flow, such as the time spent for the design (D) and validation/verification (V) stages, the number of iterations, and the complexity of design representation at each stage. The table illustrates the main steps required to create a full and validated design.  A co-design approach requires being able to have a rapid transition between each step to inform the other steps. 

The HLS model description requires approximately 1,000 lines of code. This stage is fast ($\sim$1 second) but requires several hundred iterations to optimize the algorithm performance, driven by the physics goals. The HLS stage determines the level of parallelism in the design, choice of pipelining, resource reuse factor, and clock frequency. This directly impacts the total area, power consumption, and the latency of the design. One hot encoding of finite state machines for robust Single Event Upset (SEU) prevention also needs to be introduced at this stage. Clock gating is employed to save system level power. The digital implementation stage is time intensive, requiring $\sim$65 hours of design and verification to meet the speed and area constraints with fewer iterations.

The final implementation results are presented in Table~\ref{tab:comparison}.  While there are challenges due to technology choices in making a comparison with a our design in an FPGA, we consider a fully unrolled FPGA implementation on a typical Xilinx Kintex Ultrascale FPGA device. Considering algorithm block power only and depending on configuration choices, an equivalent FPGA implementation consumes roughly 2.5--5\,W with a latency of $\sim$300\,ns~\cite{hls4ml} compared to the ASIC implementation, which consumes 95\,mW.  The ASIC implementation is expected to provide more than an order of magnitude improvement in power with a reduction in latency.

\begin{table}[tbh!]
    \begin{center}
    \caption{\label{tab:comparison}Key simulation performance parameters of the design.}
    \begin{tabular}{c|c|c|c}
        % \multicolumn{4}{*}{Key parameters} \\
        Latency & Energy/inference & Power & Area  \\
        \toprule
        $50$\,ns & $2.38$\,nJ/inf. & $95$\,mW & $3.6\,\text{mm}^2$\\
        \bottomrule
    \end{tabular}
    \end{center}
\end{table}

%%%%%%%%%%%%%%%%%%%%%%%%%%%%%%%%%%%%%% Conclusions
\section{Summary and Outlook}
\label{sec:conclusions}

A design methodology spanning from machine learning model generation to ASIC IP block creation has been presented. 
A low power, low latency reconfigurable data compression algorithm based on a convolutional neural network has been processed through synthesis and physical layout flows based on a 65\,nm LP CMOS process, designed to withstand radiation environments of up to 200\,Mrad.  

For the ECON-T ASIC, our task is to perform efficient lossy compression of an HGCAL sensor energy pattern to transmit data to off-detector electronics.  Compression is accomplished using a neural network autoencoder consisting of convolution and dense neural network layers.  
Optimal design and training of the algorithm is performed using quantization-aware training techniques to achieve good physics performance while optimizing for low power and area.  

The encoder architecture set by the model requires approximately 225,000 multiply-accumulate operations to perform vector multiplications every 25\,ns. In order to optimize for low power operation while maintaining data throughput of 40\,MHz, a highly parallel architecture is chosen at the expense of larger area. The energy consumption per inference is 2.38\,nJ. The final design consists of 800k gates and occupies a total area of 3.6\,mm$^2$.

The design demonstrates how complex NN architectures can be implemented on the front-end ASICs with realistic area constraints, allowing for minimal loss of information in the trigger data stream. Furthermore, we show that in spite of a fixed ASIC implementation, ML algorithms can still be designed with sufficient flexibility to enable reconfiguration for new operational conditions. Apart from that, the I\textsuperscript{2}C block allows real-time reconfiguration of weights, thus, facilitating the first steps toward real-time embedded AI.  This is the first time that a radiation tolerant on-detector ASIC implementation of a neural network has been designed for particle physics applications.

We look forward to inclusion of the design IP in the ECON-T ASIC fabrication. This would allow us to test the design in a physical chip. Beyond the ECON-T ASIC, there is vast potential for future work using our design methodology including: other domain applications and adaptations for ultra low power, longer latency applications; other technology nodes and design considerations; and more types of neural network architectures, which could be scaled out to larger and more complex designs.  

% {\color{red}MAYBE SAY HERE THAT IT IS CURRENTLY UNDER MANUFACTURING OR STH?}

%%%%%%%%%%%%%%%%%%%%%%%%%%%%%%%%%%%%%% FUTURE WORK
% \section{Future work}
% \label{sec:future_work}
% {\color{red} THIS IS COMPLETELY MISSING.}

%%% PREVIOUS FIGURE
\section*{Authors' Contributions}
The project was initiated and coordinated by FF, JH, and NT.  CH, MK, and DN led the development of the algorithm training and HLS translation. GDG and CH performed HLS to RTL synthesis and validation of the RTL design. FF and YL focused on the digital implementation, power analysis, and algorithm interfaces to the rest of the chip including I2C with CG. FF, MBV, and LM developed techniques for making the design radiation tolerant. JH, SOM, and NT provided supervision for the project. JD, PH, VL, JN, MP, and SS provided support for the \texttt{hls4ml} infrastructure and initial algorithm implementations in HLS.

\section*{Acknowledgment}
The authors would like to acknowledge CAD support from Sandeep Garg and Anoop Saha from Mentor Graphics for Catapult HLS and Bruce Cauble and Brent Carlson from Cadence for Innovus and Incisive. The authors would also like to thank the Fermilab ASIC group for incorporating the autoencoder block into the ECON-T ASIC; CMS HGCAL and Jean-Baptiste Sauvan for providing simulated module images for training; and Andre Davide for extensive input on network optimization.

We acknowledge the Fast Machine Learning collective as an open community of multi-domain experts and collaborators. This community was important for the development of this project.

%\vspace{-1ex}

{\footnotesize
\bibliographystyle{IEEEtran}
\bibliography{refs}

\begin{thebibliography}{10}
\providecommand{\url}[1]{#1}
\csname url@samestyle\endcsname
\providecommand{\newblock}{\relax}
\providecommand{\bibinfo}[2]{#2}
\providecommand{\BIBentrySTDinterwordspacing}{\spaceskip=0pt\relax}
\providecommand{\BIBentryALTinterwordstretchfactor}{4}
\providecommand{\BIBentryALTinterwordspacing}{\spaceskip=\fontdimen2\font plus
\BIBentryALTinterwordstretchfactor\fontdimen3\font minus
  \fontdimen4\font\relax}
\providecommand{\BIBforeignlanguage}[2]{{%
\expandafter\ifx\csname l@#1\endcsname\relax
\typeout{** WARNING: IEEEtran.bst: No hyphenation pattern has been}%
\typeout{** loaded for the language `#1'. Using the pattern for}%
\typeout{** the default language instead.}%
\else
\language=\csname l@#1\endcsname
\fi
#2}}
\providecommand{\BIBdecl}{\relax}
\BIBdecl

\bibitem{Albertsson:2018maf}
K.~Albertsson \emph{et~al.}, ``{Machine Learning in High Energy Physics
  Community White Paper},'' \emph{J. Phys. Conf. Ser.}, vol. 1085, p. 022008,
  2018.

\bibitem{Radovic:2018dip}
A.~Radovic, M.~Williams, D.~Rousseau, M.~Kagan, D.~Bonacorsi, A.~Himmel,
  A.~Aurisano, K.~Terao, and T.~Wongjirad, ``{Machine learning at the energy
  and intensity frontiers of particle physics},'' \emph{Nature}, vol. 560,
  p.~41, 2018.

\bibitem{Bourilkov:2019yoi}
D.~Bourilkov, ``{Machine and Deep Learning Applications in Particle Physics},''
  \emph{Int. J. Mod. Phys. A}, vol.~34, p. 1930019, 2020.

\bibitem{Carleo:2019ptp}
G.~Carleo, I.~Cirac, K.~Cranmer, L.~Daudet, M.~Schuld, N.~Tishby,
  L.~Vogt-Maranto, and L.~Zdeborov\'a, ``{Machine learning and the physical
  sciences},'' \emph{Rev. Mod. Phys.}, vol.~91, p. 045002, 2019.

\bibitem{hls4ml}
J.~Duarte \emph{et~al.}, ``{Fast inference of deep neural networks in FPGAs for
  particle physics},'' \emph{JINST}, vol.~13, p. P07027, 2018.

\bibitem{CMSP2L1T}
\BIBentryALTinterwordspacing
{CMS Collaboration}, ``The {Phase-2} upgrade of the {CMS} {Level-1} trigger,''
  CMS Technical Design Report CERN-LHCC-2020-004. CMS-TDR-021, 2020. [Online].
  Available: \url{https://cds.cern.ch/record/2714892}
\BIBentrySTDinterwordspacing

\bibitem{CERN-LHCC-2017-023}
\BIBentryALTinterwordspacing
------, ``The {Phase-2} upgrade of the {CMS} endcap calorimeter,'' CMS
  Technical Design Report CERN-LHCC-2017-023. CMS-TDR-019, 2017. [Online].
  Available: \url{https://cds.cern.ch/record/2293646}
\BIBentrySTDinterwordspacing

\bibitem{Collaboration_2008}
S.~Chatrchyan \emph{et~al.}, ``{The CMS Experiment at the CERN LHC},''
  \emph{JINST}, vol.~3, p. S08004, 2008.

\bibitem{cong2011high}
J.~Cong, B.~Liu, S.~Neuendorffer, J.~Noguera, K.~Vissers, and Z.~Zhang,
  ``High-level synthesis for {FPGAs}: From prototyping to deployment,''
  \emph{IEEE Trans. Comput.-Aided Des. Integr. Circuits Syst.}, vol.~30, no.~4,
  p. 473, 2011.

\bibitem{qkeras}
C.~N. Coelho, A.~Kuusela, S.~Li, H.~Zhuang, T.~Aarrestad, V.~Loncar,
  J.~Ngadiuba, M.~Pierini, A.~A. Pol, and S.~Summers, ``Automatic deep
  heterogeneous quantization of deep neural networks for ultra low-area,
  low-latency inference on the edge at particle colliders,'' 2020.

\bibitem{2020arXiv200408906K}
A.~Karbachevsky, C.~Baskin, E.~Zheltonozhskii, Y.~Yermolin, F.~Gabbay, A.~M.
  Bronstein, and A.~Mendelson, ``Early-stage neural network hardware
  performance analysis,'' \emph{Sustainability}, p. 717, 2021.

\bibitem{cmssw}
{CMS Team}, ``{CMSSW on Github}. \url{http://cms-sw.github.io/}.''

\bibitem{emd}
P.~T. Komiske, E.~M. Metodiev, and J.~Thaler, ``Metric space of collider
  events,'' \emph{Phys. Rev. Lett.}, vol. 123, p. 041801, 2019.

\bibitem{relu1}
V.~Nair and G.~E. Hinton, ``Rectified linear units improve restricted
  {Boltzmann} machines,'' in \emph{27th International Conference on
  International Conference on Machine Learning}, ser. ICML'10.\hskip 1em plus
  0.5em minus 0.4em\relax Madison, WI, USA: Omnipress, 2010, p. 807.

\bibitem{relu2}
\BIBentryALTinterwordspacing
X.~Glorot, A.~Bordes, and Y.~Bengio, ``Deep sparse rectifier neural networks,''
  in \emph{14th International Conference on Artificial Intelligence and
  Statistics}, G.~Gordon, D.~Dunson, and M.~Dudík, Eds., vol.~15.\hskip 1em
  plus 0.5em minus 0.4em\relax Fort Lauderdale, FL, USA: JMLR, 4 2011, p. 315.
  [Online]. Available: \url{http://proceedings.mlr.press/v15/glorot11a.html}
\BIBentrySTDinterwordspacing

\bibitem{vivadohls2020}
{Xilinx}, ``{Vivado High-Level Synthesis},''
  \url{https://www.xilinx.com/products/design-tools/vivado/integration/esl-design.html},
  2020.

\bibitem{catapulthls2020}
{Mentor/Siemens}, ``{Catapult High-Level Synthesis},''
  \url{https://www.mentor.com/hls-lp/catapult-high-level-synthesis}, 2020.

\bibitem{osha_resolution}
{Open Source Hardware Association}, ``{A Resolution to Redefine SPI Signal
  Names},''
  \url{https://www.oshwa.org/a-resolution-to-redefine-spi-signal-names}, 2020.

\bibitem{hlsverific2020}
{Mentor/Siemens}, ``{Catapult High-Level Synthesis - Verification},''
  \url{https://www.mentor.com/hls-lp/catapult-high-level-synthesis/hls-verification},
  2020.

\bibitem{hlslibs2020}
Mentor/Siemens, ``{HLSLibs: Open-Source High-Level Synthesis IP Libraries},''
  \url{https://hlslibs.org}, 2020.

\bibitem{alia2017lhc}
R.~G. Al{\'\i}a, M.~Brugger, F.~Cerutti, S.~Danzeca, A.~Ferrari, S.~Gilardoni,
  Y.~Kadi, M.~Kastriotou, A.~Lechner, C.~Martinella \emph{et~al.}, ``{LHC and
  HL-LHC: Present and future radiation environment in the high-luminosity
  collision points and RHA implications},'' \emph{IEEE Trans. Nucl. Sci.},
  vol.~65, p. 448, 2018.

\bibitem{huhtinen1996radiation}
M.~Huhtinen, ``The radiation environment at the cms experiment at the lhc,''
  Ph.D. dissertation, Helsinki U. Tech., 1996.

\bibitem{schrimpf2004radiation}
R.~D. Schrimpf and D.~M. Fleetwood, \emph{Radiation effects and soft errors in
  integrated circuits and electronic devices}.\hskip 1em plus 0.5em minus
  0.4em\relax World Scientific, 2004, vol.~12.

\bibitem{petersen2011single}
E.~Petersen, \emph{Single event effects in aerospace}.\hskip 1em plus 0.5em
  minus 0.4em\relax John Wiley \& Sons, 2011.

\bibitem{dodd2010current}
P.~Dodd, M.~Shaneyfelt, J.~Schwank, and J.~Felix, ``Current and future
  challenges in radiation effects on cmos electronics,'' \emph{IEEE Trans.
  Nucl. Sci.}, vol.~57, p. 1747, 2010.

\bibitem{Casas:2275135}
\BIBentryALTinterwordspacing
L.~M.~J. Casas, D.~Ceresa, S.~Kulis, S.~Miryala, J.~Christiansen, R.~Francisco,
  and D.~Gnani, ``{Characterization of radiation effects in 65 nm digital
  circuits with the DRAD digital radiation test chip},'' \emph{JINST}, vol.~12,
  no.~02, p. C02039. 11 p, 2017. [Online]. Available:
  \url{https://cds.cern.ch/record/2275135}
\BIBentrySTDinterwordspacing

\bibitem{8194929}
J.~A. {Maharrey}, J.~S. {Kauppila}, R.~C. {Harrington}, P.~{Nsengiyumva}, D.~R.
  {Ball}, T.~D. {Haeffner}, E.~X. {Zhang}, B.~L. {Bhuva}, W.~T. {Holman}, and
  L.~W. {Massengill}, ``Dual-interlocked logic for single-event transient
  mitigation,'' \emph{IEEE Trans. Nucl. Sci.}, vol.~65, no.~8, pp. 1872--1878,
  2018.

\bibitem{fulkerson2006single}
D.~Fulkerson, ``Single-event-effect hardened circuitry,'' Nov.~30 2006, uS
  Patent App. 11/136,920.

\bibitem{kuboyama2009single}
S.~Kuboyama, H.~Shindou, Y.~Iide, and A.~Makihara, ``Single-event-effect
  tolerant soi-based inverter, nand element, nor element, semiconductor memory
  device and data latch circuit,'' 2009, uS Patent 7,504,850.

\bibitem{gong2008new}
R.~Gong, W.~Chen, F.~Liu, K.~Dai, and Z.~Wang, ``A new approach to single event
  effect tolerance based on asynchronous circuit technique,'' \emph{JETTA},
  vol.~24, p.~57, 2008.

\bibitem{lyons1962use}
R.~E. Lyons and W.~Vanderkulk, ``The use of triple-modular redundancy to
  improve computer reliability,'' \emph{IBM J. Res. Dev}, vol.~6, p. 200, 1962.

\bibitem{habinc2002functional}
S.~Habinc, ``Functional triple modular redundancy (ftmr). vhdl design
  methodology for redundancy in combinatorial and sequential logic,''
  \emph{Gaisler Research, Design and Assessment Report (Version 0.2)}, 2002.

\end{thebibliography}
}

\end{document}